\documentclass{article}

\usepackage[utf8]{inputenc}
\usepackage{tismir}
\usepackage{amsmath}
\usepackage{hyperref}
\usepackage{url}
\usepackage{graphicx}
\usepackage{booktabs}

\usepackage{microtype}
\usepackage{paralist}
\usepackage{fancybox}
\usepackage{units}
\usepackage{pdflscape}

\title{An Interdisciplinary Review of Music Performance Analysis}

\author{%
Alexander Lerch\thanks{Center for Music Technology, Georgia Institute of Technology},%
~Claire Arthur\protect\footnotemark[1],%
~Ashis Pati\protect\footnotemark[1],%
~and Siddharth Gururani\protect\footnotemark[1]}

\date{}
\authorref{Lerch, A., Arthur, C., Pati, A., Gururani, S.}
\authorshort{Lerch et al.} 

\begin{document}

\twocolumn[{%
\maketitleblock
\begin{abstract}
	A musical performance renders an acoustic realization of a musical score or other representation of a composition. Different performances of the same composition may vary in terms of various performance parameters such as timing or dynamics, and these variations may have a major impact on how a listener perceives the music. The analysis of music performance has traditionally been a peripheral topic for the MIR research community, where often a single audio recording is used as representative of a musical work. This paper surveys the field of Music Performance Analysis (MPA) from several perspectives including the measurement of performance parameters, the relation of those parameters to the actions and intentions of a performer or perceptual effects on a listener, and finally the assessment of musical performance. This paper also discusses MPA as it relates to MIR, pointing out opportunities for collaboration and future research in both areas.
\end{abstract}
}]
\saythanks{}
\section{Introduction}\label{sec:introduction}
    Music, as a performing art, requires a performer or group of performers to render a musical ``blueprint'' into an acoustic realization \citep{hill_score_2002, clarke_understanding_2002}. {This musical blueprint can, for example, be a score as in the case of Western classical music, a lead sheet for jazz, or some other genre-dependent representation describing the compositional content of a piece of music. The performers are usually musicians but might also be, e.g., a computer rendering audio.}
    
    The performance plays a major role in how listeners perceive a piece of music: even if the blueprint is identical for different renditions, as is the case in
    Western classical music, listeners may prefer one performance over another and appreciate different `interpretations' of the same piece of music. These differences
    are the result of the performers' {intentionally or unintentionally} interpreting, modifying, adding to, and dismissing information from the score or blueprint {(for the sake of simplicity, the remainder of this text will use the terms score and blueprint synonymously)}. This constant re-interpretation of music is inherent to the art form and is a vital and expected component of music.

\begin{figure*}%
        \centering
        \includegraphics[width=\linewidth]{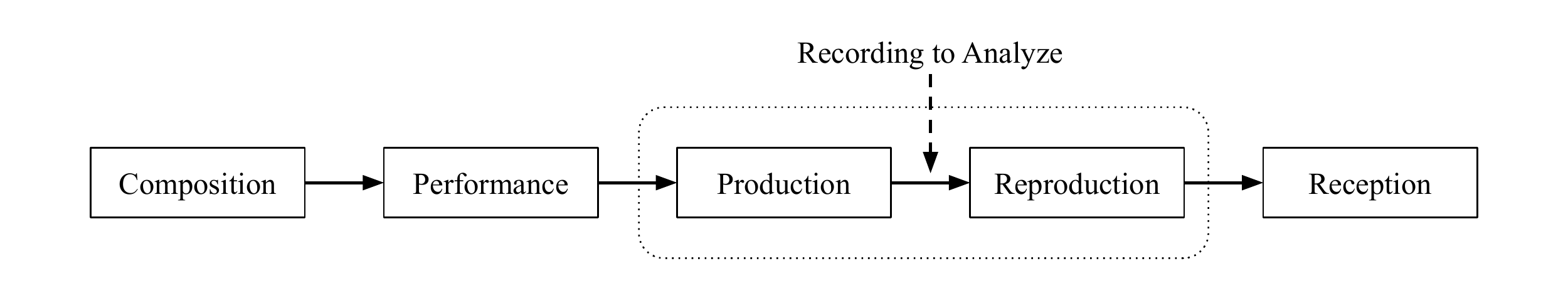}%
        \caption{Chain of musical communication from composer to listener. The dotted box disappears for performances that are not recorded.}%
        \label{fig:chain}%
    \end{figure*}

    Formally, musical communication can be described as a chain as shown in Figure~\ref{fig:chain}: the composer {typically} only communicates with the listener through the performer who renders the blueprint to convey musical ideas to the listener \citep{kendall_communication_1990}. 
    The performer does this by varying musical parameters while leaving the compositional content untouched.
    The visualized model of the communication chain displays a one-way communication path of information transmission. There can be, however, flows of information in the other direction, influencing the performance itself.
    Such a feedback path might transport information such as the instrument's sound and vibration \citep{todd_vestibular_1993}, the room acoustics, \citep{luizard_singing_2020}, and the audience reaction \citep{kawase_importance_2014}. 
    Although the recording studio lacks an audience, a performance can also evolve during a recording session. \citet{katz_capturing_2004} points out that in such a session, performers will listen to the recording of themselves and adjust ``aspects of style and interpretation.'' In addition, the producer might also have impact on the recorded performance \citep{maempel_musikaufnahmen_2011}.
    \newline\indent
    The large variety of performance scenarios makes it necessary to focus on the common core of all music performances: the audio signal. While a music performance
    can, for example, contain visual information such as gestures and facial expressions \citep{bergeron_hearing_2009,platz_when_2012,tsay_sight_2013}, not every music performance has these cues. A musical robot, for
    example, may or may not convey such cues. The acoustic rendition, however, is the integral part of a music performance that cannot be missing; simply put, there
    exists no music without sound. An audio recording is a fitting representation of the sound that allows for quantitative analysis. This audio-focused view should
    neither imply that non-audio information cannot be an important part of a performance nor that non-audio information cannot be analyzed. 
    The focus on the \textit{audio recording} of a performance makes it important to recognize that every recording contains processing choices and interventions by the
    production team with potential impact on the expressivity of the recording. \citet{maempel_musikaufnahmen_2011} discusses as main influences in the context of classical music the sound engineer (dynamics, timbre, panorama, depth) and the editor (splicing of different tracks, tempo and timing, pitch). These manipulations and any restrictions of the recording and distribution medium are part of the audio to be analyzed and cannot be separated anymore from the performers' creation. As the release of a recording usually has to be pre-approved by the performers it is assumed that the performers' intent has not been distorted.
    
    \begin{figure}[t]
        \centerline{
        \includegraphics[width=\columnwidth]{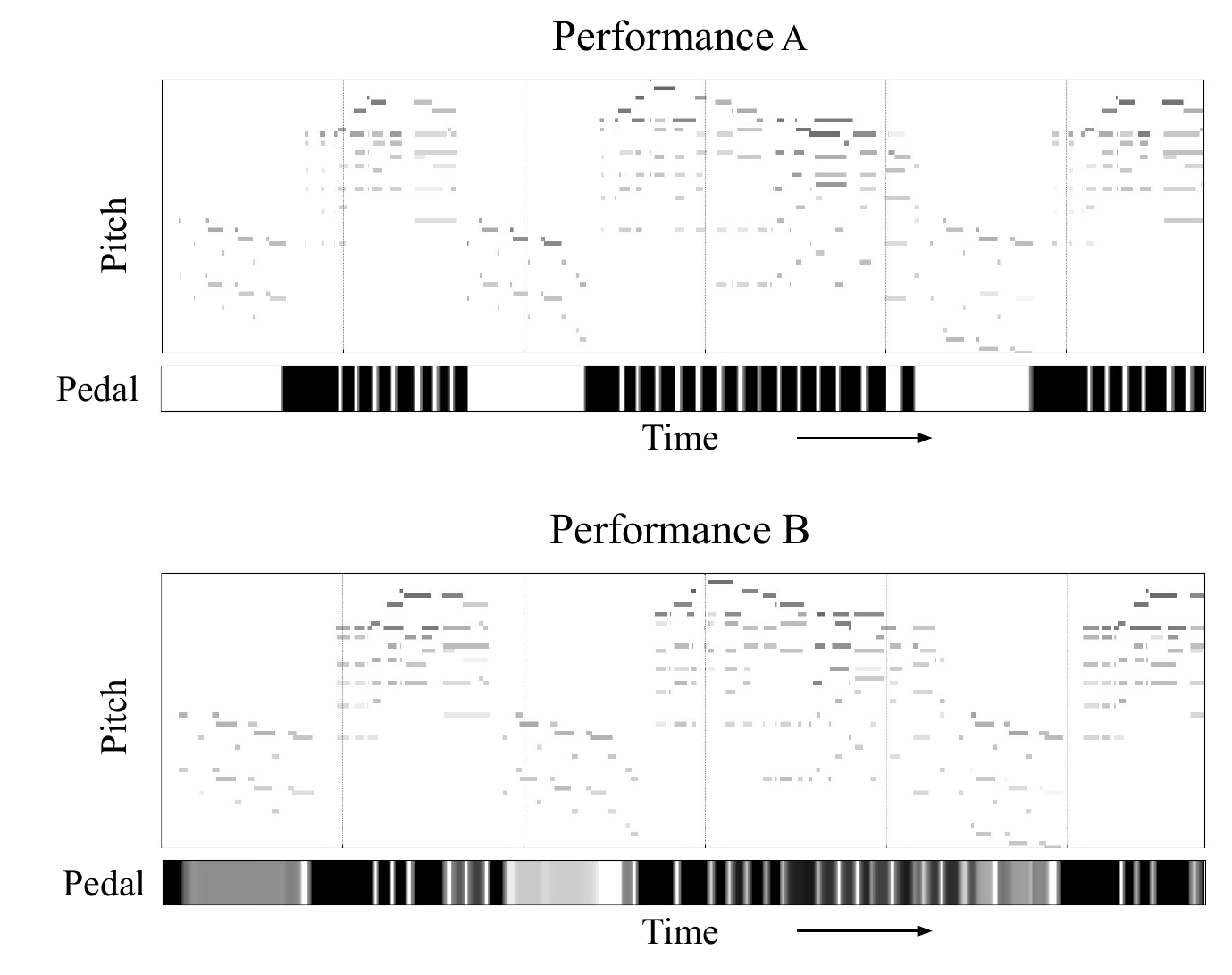}}
        \caption{Variations in two different performances of Frédéric Chopin's Fantasie in F Minor, Op.~49 (taken from the Maestro dataset \citep{hawthorne_enabling_2019}. For each performance, timing and dynamics are shown using the piano rolls (darker color indicates higher velocity). Pedal control is shown below the piano roll (darker color indicates increasing usage of the pedal). Visualized excerpts correspond to the beginning of the piece.}
    \label{fig:mpa_performance_viz}
    \end{figure}
   Although the change of performance parameters can have a major impact on a listener's perception of the music \citep[also Section~\ref{sec:listener}]{clarke_listening_2002}, the nature of the performance parameter variation is often subtle, and must be evaluated in reference to something; 
typically either the same performer and piece at a different time, different performances of the same piece, or deviations from a perfectly quantized performance rendered from a symbolic score representation (e.g., MIDI) without tempo or dynamics variations.
Notice that this reference problem makes the genre of music for performance analysis biased towards classical music, as there is commonly an abstract (i.e., score) representation that easily fills the model of the quantized version of a piece of music.
In addition, the classical model of performance is based on a finite number of pre-composed musical works performed by numerous individuals over time, thus making it a rich resource for performance analysis.  
Figure~\ref{fig:mpa_performance_viz} visualizes two example piano performances of the beginning of Frédéric Chopin's \textit{Fantasie in F Minor, Op.~49}, where variations in local tempo, dynamics, and pedaling can easily be identified.
Even in musical traditions that also have scores, such as jazz, the role of the performer as `interpreter' of the score is complicated by normative practices such as
elaborate embellishment and improvisation (often to the point of highly obscuring the original material). Other genres might completely eliminate the concept of separable composition and performance. \citet{mcneil_seed_2017} argues, for example, that the distinction between `composition' and `improvisation' cannot capture the essence of performance creativity in Indian Hindustani music, where the definition of fixed `composed' material and improvised material becomes hard and it might be more meaningful to refer to seed ideas which grow and expand throughout the performance.
    
    Although the distinction between score and performance parameters is less obvious for non-classical genres of Western music, especially ones without clear separation between the composer and the performer, the concept and role of a performer as interpreter of a composition is still very much present, be it as a live interpretation of a studio recording or a cover version of another artist's song. In these cases, the freedom of the performers in modifying the score information is often much higher than it is for classical music~---~reinterpreting a jazz standard can, for example, include the modification of content related to pitch, harmony, and rhythm.
    
    Formally, performance parameters can be structured in the same basic categories that we use to describe musical audio in general: tempo and timing, dynamics, pitch, and timbre \citep{lerch_introduction_2012}. While the importance of different parameters might vary from genre to genre, the following list introduces some mostly genre-agnostic examples to clarify these performance parameter categories: %
    \begin{compactitem}
        \item   
        \textit{Tempo and Timing}: the score specifies the general rhythmic content, just as it often contains a tempo indicator. While the rhythm is often only slightly modified by performers in terms of micro-timing, the tempo (both in terms of overall tempo as well as expressive tempo variation) is frequently seen only as a suggestion to the performer.
        \item   
        \textit{Dynamics}: in most cases, score information on dynamics is missing or only roughly defined. The performers will vary loudness and highlight specific events with accents based on their plan for phrasing and tension, and the importance of certain parts of the score. 
        \item   
        \textit{Pitch}: the score usually defines the general pitches to play, but pitch-based performance parameters include expressive techniques such as vibrato as well as {intentional or unintentional} choices for intonation.
        \item   
        \textit{Timbre}: as the least specific category of musical parameters, scores encode timbre parameters often only implicitly (e.g., instrumentation) while performers can, for example, change playing techniques or the choice of specific instrument configurations (such as the choice of organ registers).
    \end{compactitem}
There exist many performance parameters and playing techniques that either cannot be easily associated with one of the above categories or span multiple categories; examples of such parameters are articulation (\textsl{legato}, \textsl{staccato}, \textsl{pizzicato}) or forms of ornamentation in Baroque or jazz music.
   
    One intuitive form of Music Performance Analysis (MPA)~---discussing, criticizing, and assessing a performance after a concert---~has arguably taken place
    since music was first performed. Traditionally, however, such reviews are qualitative and not empirical.
    While there are a multitude of approaches to music performance analysis, and numerous factors shape the insights and outcomes of such analyses, in this literature review we are only concerned with quantitative, systematic approaches to MPA scholarship.
    However, it is easily acknowledged that the design of any empirical study and the interpretation of results could be improved by a careful consideration of the musicological context (e.g., the choice of edition that a classical performance is rendered from~---~see \cite{rink_respect_2003}). 
    Nevertheless, a detailed discussion of such contexts is beyond the scope of this article.

    Early attempts at systematic and empirical MPA can be traced back to the 1930s with vibrato and singing analysis by \citet{seashore_psychology_1938} and the examination of piano rolls by \citet{hartmann_untersuchungen_1932}. In the past two decades, MPA has greatly
    benefited from the advances in audio analysis made by members of the Music Information Retrieval (MIR) community, significantly extending the volume of empirical data by simplifying access to a continuously growing heritage of commercial audio recordings.
    While advances in audio content analysis have had clear impact on MPA, the opposite is less true. An informal search reveals that while there have been
    publications on performance analysis at ISMIR, the major MIR conference, their absolute number remains comparably small \citep[compare][]{toyoda_utility_2004,takeda_rhythm_2004,chuan_dynamic_2007,sapp_comparative_2007,sapp_hybrid_2008,hashida_new_2008,liem_expressive_2011,okumura_stochastic_2011,devaney_study_2012,jure_pitch_2012,van_herwaarden_predicting_2014,liem_comparative_2015,arzt_real-time_2015,page_toolkit_2015,xia_spectral_2015,bantula_jazz_2016,peperkamp_formalization_2017,gadermaier_study_2019,maezawa_rendering_2019}
    with a title referring to ``music performance'' out of approximately $1,950$ ISMIR papers overall).

    Historically, MIR researchers often do not distinguish between score-like information and performance information even if the research deals with audio recordings of performances. For instance, the goal of music transcription, a very popular MIR task, is usually to transcribe all pitches with their onset times \citep{benetos_automatic_2013}; that means that a successful transcription system transcribes two renditions of the same piece of music differently, although the ultimate goal is to detect the same score (note that this is not necessarily true for all genres).
    Therefore, we can identify a disconnect between MIR research and performance research that impedes both the evolution of MPA approaches and robust MIR algorithms, slows gaining new insights into music aesthetics, and hampers the development of practical applications such as new educational tools for music practice and assessment. This paper aims at narrowing this gap by introducing and discussing MPA and its challenges from an MIR perspective. In pursuit of this goal, this paper complements previous review articles on music performance research \citep{sloboda_music_1982,palmer_music_1997, gabrielsson_performance_1999, gabrielsson_music_2003, goebl_sense_2005} and expands on \citet{lerch_music_2019} by 
    \begin{inparaitem}[]
        \item   integrating non-classical and non-Western music genres,
        \item   including a more extensive number of relevant publications, and
        \item   clearly outlining the challenges music performance research is facing.
    \end{inparaitem}
Note that while performance research has been inclusive of various musical genres, such as the Jingju music of the Beijing opera \citep{zhang_understanding_2017,gong_automatic_2018}, traditional Indian music \citep{clayton_time_2008,gupta_objective_2012,narang_acoustic_2017} and jazz music \citep{abeser_score-informed_2017}, the vast majority of studies are concerned with Western classical music. The focus on Western music can also be observed in the field of MIR in general, despite efforts in diversifying the field \citep{serra_creating_2014,tzanetakis_computational_2014}. As mentioned, the reason for the focus on classical music within MPA may be due to the clear systematic differentiation between score and performance. This imbalance means that this overview article will necessarily emphasize Western classical music while referring to other musical styles wherever appropriate.
 
The remainder of this paper is structured as follows. The following Section~\ref{sec:descript} presents research on the objective description, modeling, and visualization of the performance itself, identifying commonalities and differences
between performances. The subsequent sections focus on studies taking these objective performance parameters and relating them to either the performer
(Section~\ref{sec:performer}), the listener (Section~\ref{sec:listener}), or the assessment of the performer from a listener's perspective (Section~\ref{sec:assessment}). We conclude our overview with a summary on applications of MPA and final remarks in Section~\ref{sec:apps}. 

\section{{Performance Measurement}}\label{sec:descript}
    A large body of work focuses on an exploratory approach to analyzing performance recordings and describing performance characteristics. 
Such studies typically extract characteristics such as the tempo curve or histogram \citep{repp_patterns_1990, palmer_mapping_1989, povel_temporal_1977,srinivasamurthy_aspects_2017} or loudness curve \citep{repp_microcosm_1998, seashore_psychology_1938} from the audio and aim at either gaining general knowledge on performances or comparing attributes between different performances/performers based on trends observed in the extracted data. Additionally, there are also studies focusing on discovery of general patterns in performance parameters, which can be useful in identifying trends such as changes over eras \citep{ornoy_analysis_2018}.

\subsection{Tempo, timing, and dynamics}
    Deviations in tempo, timing and dynamics are considered to be some of the most salient performance parameters and hence have been the focus of various studies in MPA. While these performance parameters have been studied in isolation in some instances, we present them together since there is substantial work aiming to understand their interrelation. 

Close relationships were observed between musical phrase structure and deviations in tempo and timing \citep{povel_temporal_1977, shaffer_timing_1984, palmer_music_1997}. 
For example, tempo changes in the form of \textsl{ritardandi} tend to occur at phrase boundaries \citep{palmer_mapping_1989,lerch_software-based_2009}. As a related structural cue, \citet{chew_playing_2016} proposed the concept of tipping points in the score, leading to a timing deviation with extreme pulse variability in the context of Western classical music performance.

Correlations were observed between timing and dynamics patterns \citep{repp_dynamics_1996,lerch_software-based_2009}. 
\citet{dalla_bella_tempo_2004} found the overall tempo influences the overall loudness of a performance. 
There are also indications that loudness can be linked to pitch height \citep{repp_dynamics_1996}. 
\citet{cheng_quantitative_2008} analyzed global phrasing strategies for violin performance using loudness and tempo variation profiles and found dynamics to be more closely related to phrasing than tempo.
While the close relation of tempo and dynamics to structure has been repeatedly verified, \citet{lerch_software-based_2009} did not succeed in finding similar relationships between structure and timbre properties in the case of string quartet recordings.

In the context of jazz performances, \citet{wesolowski_timing_2016} found that both score and performance parameters such as underlying harmony, pitch interval size, articulation, and tempo had significant correlations with timing variations between successive eighth notes. He also found these parameters correlated with synchronicity between separate parts of a jazz ensemble. 
Several researchers conducted experiments studying swing style jazz \citep{ellis_analysis_1991,progler_searching_1995,collier_study_2002,friberg_swing_2002}. Such studies focused on measuring or quantifying discrepancies and asynchrony of performers in order to study what characteristics of jazz performances made them `swing' \citep{friberg_swing_2002}. \citet{ellis_analysis_1991} found that asynchrony 
of jazz swing performers to the prevailing meter is positively correlated with the tempo and consists mainly of delaying attacks. \citet{progler_searching_1995} noted that participatory asynchrony in swing is observed and measurable at a subsyntax level.
\citet{abeser_dynamics_2014} studied the relationship of note dynamics in jazz improvisation with other contextual information such as note duration, pitch, and position in the score. They used a score-informed music source separation algorithm to isolate the solo instrument and found that higher and longer notes tend to be louder, and structural accents are typically emphasized.
\citet{busse_toward_2002} conducted an experiment to objectively measure deviations in terms of timing, articulation, and dynamics of jazz swing performers from mechanical regularity using MIDI-based `groove quantization'. 
He created reference performer models using the measured performance parameter deviations and compared them against mechanical or quantized models. Experts were asked to rate the `swing representativeness' of the different models. 
He found that reference performer models were rated to be representative of swing whilst similar mechanical models of performance were rated poorly.
\citet{ashley_nt_2002} described timing in jazz ballad performance as melodic rhythm flexibility over a strict underlying beat pattern, which is a type of rubato. 
He found timing deviations to have strong relationships with musical structure.
\citet{collier_exploration_1994} studied tempo in corpora of jazz performances. They manually timed these recordings and found that tempo was normally {(Gaussian)} distributed when computed in terms of metronome markings but not when computed using note durations. They noted that while jazz performances are stable in terms of tempo, systematic patterns in timing variabilities tend to serve expressive functions. {\citet{iyer_embodied_2002} contrasted micro-timing in African-American music from {techniques in Western}  classical music such as \textsl{rubato} and \textsl{ritardando}. Employing an embodied cognition framework, he argued that the differences are due to the emphasis of the human body in the cultural aesthetics of African-American musics.}

Studies analyzing musical timing are not only limited to Western music. \citet{srinivasamurthy_aspects_2017} performed a relatively large-scale computational analysis of rhythm in Hindustani classical percussion, confirming and quantifying tendencies pertaining to timing such as deviations of tempo within a metric cycle {(also referred to as \textit{tal})}.
The study demonstrated the value of using MIR techniques for rhythm analysis of large corpora of music.
\mbox{\citet{bektas_relationships_2005}} confirmed the relationship between prosodic meter \textit{arûz} and musical meter \textit{usûl} in Turkish vocal music and found the existence of an even stronger concordance between a subset of prosodic patterns called \textit{bahir} and \textit{usûl}.

In addition to the studies already presented, there is a large body of work that focuses on modeling of timing, tempo, and dynamics. Section~\ref{subsec:modeling} discusses such modeling approaches to performance measurement in detail.

\subsection{Tonal analysis}

Pitch-based performance parameters have been analyzed mostly in the context of single-voiced instruments. For instance, the vibrato range and rate has been studied for vocalists \citep{seashore_psychology_1938,devaney_automatically_2011} and violinists \citep{bowman_macleod_influences_2006, dimov_short_2010}. Regarding intonation, \citet{devaney_automatically_2011} found significant differences between professional and non-professional vocalists in terms of the size of the interval between semi-tones.

Studies have also been conducted on the relationship between pitch and meter in the context of bebop style jazz \citep{jarvinen_tonal_1995, jarvinen_effect_2000}. These studies found that measurements of chorus-level tonal hierarchies match quite closely to rating profiles of chromatic pitches found in European art music and that metrical structure plays a role in determining which pitches are emphasized or de-emphasized. The studies also indicated that there is no effect of syncopation or polyrhythm on the use of certain pitches.

\citet{franz_markov_1998} demonstrated the utility of Markov chains in the analysis of jazz improvisation. He found Markov chains useful in quantifying the frequency of
notes and patterns of notes which is in turn useful as comparison tools for musical scale analyses. Furthermore, he noted that this modeling technique can be useful for stylistic comparison as well as in developing metrics for style and creativity.
\citet{frieler_midlevel_2016} proposed an analysis framework for jazz improvisation based on so-called `midlevel units' (MLU). These are musical units on the middle level between individual notes and larger form parts. They hypothesize that MLUs correspond to the improvisers' playing ideas and musical ideas and propose a taxonomy system for MLUs. The authors subsequently study the distribution of occurrences and durations of MLUs in a large corpus of jazz improvisations. They note that the most common MLUs used in improvisations belong to the \textsl{lick} and \textsl{line} categories. They also find that the distribution of MLU types differs between performers and styles.

Research on jazz solos has utilized MIR methods for source separation and pitch tracking to analyze intonation and other tonal features \citep{abeser_score-informed_2014,abeser_score-informed_2015}. 
\citet{abeser_score-informed_2014} proposed a score-informed pitch tracking algorithm for analysis. They analyze distributions of various pitch contour features for different artists to identify patterns between different artists and instruments. Several contextual parameters such as relative position of notes in a phrase, beat position in the bar, etc., are also extracted and correlations between these parameters and pitch features are studied. They found statistically significant correlations between the pitch contour features and contextual features but most of the correlations had small effect size. 
\citet{abeser_score-informed_2015} utilized score-based source separation and pitch tracking to study intonation in jazz brass and woodwind solos to identify trends in tuning frequency over various decades, intonation for different artists, and properties of vibrato in different contexts and performers.

A survey of computational methods utilized to study tonality in Turkish \textit{Makam} music was conducted by \citet{bozkurt_computational_2014}, including research related to the analysis of tuning frequency and melodic phrases, transcription of performances, \textit{Makam} recognition, as well as rhythmic and timbral analysis in \textit{Makam} music.
\citet{atli_method_2015} discuss the importance of tonic frequency or \textit{karar} estimation for Turkish \textit{Makam} music and devise a simple yet well performing method
for the task. They find that detecting the last note of a performance recording and estimating the frequency works well for \textit{karar} estimation. 
\citet{hakan_characterization_2012} analyzed Turkish \textit{ney} performances to identify key aspects of embellishments. They found that the rate of change of vibrato and `pitch bump', which measures the deviation of pitch just before ascending or descending into the next note, were the key features useful for distinguishing performance styles across various performers. 

Similarly, several researchers have worked on analysis of melody and tonality in Indian classical music. 
\citet{ganguli_towards_2017} modeled ungrammatical phrases, i.e., phrases straying away from the predetermined \textit{raga}, in Hindustani music performance. They utilized computational techniques to model the tonal hierarchies and melodic shapes of different \textit{ragas} toward that end.
\citet{viraraghavan_statistical_2017} analyzed the use of ornamentation known as \textit{gamakas} in \textit{Carnatic} music performances. They find that the use of \textit{gamakas} is vital in defining the \textit{raga} during a performance.

\citet{chen_characterization_2013} developed methods for analysis of intonation in Beijing opera or \textit{Jingju} music. The methods, involving peak distribution analysis of pitch histograms, validated claims in literature that the fourth degree is higher, and the seventh degree is lower than the corresponding pitches in the equally tempered scale. Chen also found pitch histograms to be good features to distinguish role types in opera performances.
\citet{caro_repetto_comparision_2015} utilized MIR techniques for pitch tracking and audio feature extraction to compare singing styles of two \textit{Jingju} schools, namely the \textit{Mei} and \textit{Cheng} schools. Their experiments quantitatively support observations made in musicological texts about characteristics such as pitch register,
vibrato, volume/dynamics, and timbre brightness. In addition, they find other properties not previously reported; for example, vibrato in \textit{Mei} tends to be slower and wider on average than in \textit{Cheng}. 
    \citet{yang_vibrato_2015} developed methods based on filter diagonalization and hidden Markov models to detect and model vibrato and portamento in performances of \textit{erhu}, violin and Beijing opera vocals.
There are studies aiming to understand the relationship between linguistic tone and melodic pitch contours in \textit{Jingju} music by utilizing machine learning methods such as clustering \citep{zhang_predicting_2015, zhang_study_2014}. \textit{Jingju} music utilizes a two-dialect tone system making the tone-melody relationship complicated. 

\subsection{Modeling}
\label{subsec:modeling}
    While most of the studies mentioned above make use of statistical methods to extract, summarize, visualize, and investigate patterns in the performance data, researchers have also investigated modeling approaches to better understand performances. Several overview articles exist covering research on generating expressive musical performance \citep{cancino-chacon_computational_2018,kirke_guide_2013,widmer_computational_2004}. 
    In this subsection, however, we primarily focus on methods that model performance parameters leading to useful insights, while ignoring the generative aspect.

    Several researchers have attempted to model timing variations in performances. 
    In early work, \citet{todd_kinematics_1995} modeled \textsl{ritardandi} and \textsl{accelerandi} using kinematics theory, concluding that tempo variation is analogous to velocity.
    \citet{li_clustering_2017} introduced an approach invariant to phrase length for analyzing expressive timing. They utilize Gaussian mixture models (GMMs) to model the polynomial regression coefficients for tempo curves instead of directly modeling expressive timing.
    \citet{liem_expressive_2011} proposed an entropy-based deviation measure for quantifying timing in piano performances and found it to be a good alternative to standard-deviation-based measures.
    \citet{grachten_temporal_2017} utilized recurrent neural networks to model timing in performances and demonstrated the benefits over static models that do not account for temporal dependencies between score features.
    \citet{stowell_maximum_2013} introduced a Bayesian model of tempo modulation at various time-scales in performances.

    For dynamics modeling,
    \citet{kosta_change-point_2015} applied and compared two change-point algorithms to detect dynamics changes in performed music, evaluating them using the corresponding dynamics markers in the score.
    In further research, \citet{kosta_dynamics_2018} quantified the relationships between notated and performed dynamics using a corpus of performed Chopin Mazurkas. \citet{kosta_mapping_2016} applied various machine learning methods such as decision trees, support vector machines (SVM) and neural networks to understand the relationship between dynamics markings and performed loudness. They find that score-based features are more important than performer style features for predicting dynamics markings given performed loudness and vice versa.
    Similarly, \citet{marchini_sense_2014} utilized decision trees, SVMs and k-nearest neighbor classifiers to model and predict performance features such as intensity, timing deviations, vibrato extent and bowing speed of each note in string quartet performances. They found that inter-voice attributes played a strong role in models trained with ensemble recordings versus solo recordings.
    \citet{grachten_linear_2012} introduce a so-called linear basis function model which encodes score information using weighted combinations of a set of basis functions. They utilize this model to predict and analyze dynamics in performance. \citet{grachten_temporal_2017} extend the framework with a recurrent model which better captures temporal relationships in order to improve modeling of timing. \citet{cancino-chacon_evaluation_2017} performed a large-scale evaluation of linear, non-linear and temporal models for dynamics in piano and orchestral performances. These models utilize various features extracted from the score to encode structure in dynamics, pitch and rhythm.

    An alternative direction in performance modeling research involves methods to discover rules for performance \citep{widmer_discovering_2003}. Widmer's ensemble learning method succeeds in finding simple, and in some cases novel, rules for music performance. An approach to modeling expressive jazz performance based on genetic algorithms was proposed by \citet{ramirez_genetic_2008}, learning rules to generate sequences that are best able to fit the training data.

\subsection{Visualization}

Many traditional approaches to performance parameter visualization such as pitch contours \citep{gupta_objective_2012,abeser_score-informed_2014}, tempo curves \citep{repp_patterns_1990, palmer_mapping_1989, povel_temporal_1977}, and scatter plots \citep{lerch_software-based_2009} are not necessarily interpretable or easily utilized for comparative studies. This led researchers to develop other, potentially more intuitive or condensed forms of visualization that allow describing and comparing different performances.
The `performance worm', for example, is a pseudo-3D visualization of the tempo-loudness space that allows the identification of specific gestures in that space \citep{langner_representing_2002,dixon_performance_2002}. 
\citet{sapp_comparative_2007,sapp_hybrid_2008} proposed the so-called `timescapes' and `dynascapes' to visualize subsequent similarities of performance parameters.

The `Phenicx' project explored various ways of visualizing orchestral music information, including both score and performance information \citep{gasser_classical_2015}. Dynagrams and tempograms, for example, are used to visualize various temporal levels of loudness and tempo variations, respectively.
    \citet{dittmar_swingogram_2018} devised swingogram representations by analyzing and tracking the swing ratio implied by the ride cymbal in jazz swing performance. This visualization enables insights into jazz improvisation such as the interaction between a soloist and the drummer.

\subsection{Challenges}

The studies presented in this section often follow an exploratory approach; extracting various parameters in order to identify commonalities or differences between performances. 
While this is, of course, of considerable interest, one of the main challenges is the interpretability of these results. 
Just because there is a timing difference between two performances does not necessarily mean that this difference is musically or perceptually meaningful. 
Without this link, however, results can only provide limited insights into which parameters and parameter variations are ultimately important. 

The acquisition of data for analysis is another challenge in MPA research. \citet{goebl_sense_2005} discuss various methods that have been used to that end, including special instruments (such as Yamaha Disklaviers, piano rolls), hand measurement of performance parameters, as well as automatic audio analysis tools. All these methods have potential downsides. For example, the use of special instruments and sensors excludes the analysis of performances not recorded on these specific devices and the associated formats (e.g., MIDI) may have difficulties representing special playing techniques. Manual annotation can be time consuming and tedious. The fact that the majority of studies surveyed here rely on manually annotated data implies that available algorithms for automatic performance parameter extraction lack the reliability and/or accuracy for practical MPA tasks. This is especially true for ensemble performances where the polyphonic and poly-timbral nature as well as timing fluctuations between individual voices complicate the analysis.
As a result of these challenges, most studies are performed on manually annotated data with small sample sizes, possibly leading to poor generalizability of the results. The increasing number of datasets providing performance data as listed in Table~\ref{table:dataset}, however, gives hope that this ceases to be an issue in the future.

\section{{Performer}}\label{sec:performer}
    While most studies focus on the extraction of performance parameters or the mapping of these parameters to the listeners' perception (see Sections~\ref{sec:listener} and \ref{sec:assessment}), some investigate the capabilities, goals, and strategies of performers. 
    A performance is usually based on an explicit or implicit performance plan with clear intentions \citep{clarke_understanding_2002}. {This seems to be the case also for improvised music: for instance, \citet{dean_generative_2014} could verify clearly perceivable structural boundaries in free jazz piano improvisation.} There is, as Palmer verified, a clear relation between reported intentions and objective parameters related to phrasing and timing of the performance \citep{palmer_mapping_1989}. 
    Similar relations between the intended emotionality and loudness and timing measures were reported in multiple studies \citep{juslin_cue_2000,dillon_extracting_2001,dillon_statistical_2003,dillon_recognition_2004}. For example, projected emotions such as anger and sadness show significant correlations with high and low tempo and high and low overall sound level, respectively.
    Moreover, a performer's control of expressive variation has been shown to significantly improve the conveyance of emotion. For instance,  a  study  by  \citet{vieillard_expressiveness_2012} found that listeners were better able to perceive the presence  of  specific  emotions in  music  when  the  performer played an `expressive' (versus mechanical) rendition of the composition. 
    This suggests that the performer plays a fairly large role in communicating an emotional `message' above and beyond what is communicated through the score alone \citep{juslin_communication_2003}.
In music performed from a score in particular, the score-based representation might be thought of as a set of instructions in the sense that the notational system itself is used to communicate basic structural information to the performer. 
However, as noted by \citet{rink_respect_2003}, the performer is not simply a medium or vessel through which performance directions are carried out, but ``what
performers do has the potential to impart meaning and create structural understanding.''

Research by \citet{friberg_swing_2002} set out to tackle the question of what makes music `swing.' Their approach was to examine the variation in the `swing ratio' between pairs of eighth notes in jazz music, and they found that it tends to vary as a function of tempo. 
This finding has interesting implications for MPA as well as perceptual experiments. 
Across performers there is a clear systematic relation between the stretching of the ratio at slower tempi and the compressing of the ratio at higher tempi, such that it approaches a 1:1 ratio at approximately \unit[300]{BPM}. 
Interestingly, the duration of the second eighth note remained fairly constant at approximately \unit[100]{ms} for medium to fast tempi, suggesting a practical limit on tone duration that, as the authors speculate, could be due to perceptual factors. 

    Another interesting area of research is performer error. \citet{repp_art_1996} analyzed performers' mistakes and found that errors were concentrated in mostly unimportant parts of the
    score (e.g., middle voices) where they are harder to recognize \citep{huron_tone_2001}, suggesting that performers {intentionally or unintentionally} avoid salient mistakes. 

    \subsection{Influences}
        In addition to the performance plan itself, there are other influences shaping the performance. Acoustic parameters of concert halls such as the early decay time have been shown to impact performance parameters such as tempo \citep{scharer_kalkandjiev_influence_2013,scharer_kalkandjiev_influence_2015,luizard_singing_2019,luizard_singing_2020}. 
		Related work by Repp showed that pedaling characteristics in piano performance are dependent on the overall tempo  \citep{repp_pedal_1996, repp_effect_1997}.

        Other studies investigate the importance of the feedback of the music instrument to the performer \citep{sloboda_music_1982}; there have been studies reporting on the effect of deprivation of auditory feedback \citep{repp_effects_1999}, investigating the performers' reaction to delayed or changed auditory feedback \citep{pfordresher_effects_2002, finney_auditory_2003, pfordresher_auditory_2005}, or evaluating the role of tactile feedback in a piano performance \citep{goebl_tactile_2008}. In summary, the different forms of feedback have been found to have small but significant impact on reproduction accuracy of performance parameters.

    \subsection{Challenges}
        There is a wealth of information about performances that can be learned from performers. 
        The main challenge of this direction of inquiry is that such studies have to involve the performers themselves. 
        This limits the amount of available data and usually excludes well-known and famous artists, resulting in a possible lack of generalizability.
        Depending on the experimental design, the separation of possible confounding variables (for example, motor skills, random variations, and the influence of common performance rules) from the scrutinized performance data can be a considerable challenge.

\section{{Listener}}\label{sec:listener}
    Every performance will ultimately be heard and processed by a listener. 
The listener's meaningful interpretation of the incoming musical information depends on a
sophisticated network of parameters. 
These parameters include both objective (or, at least, measurable) features that can be estimated from a score or derived from a
performance, as well as subjective and `internal' ones such as factors shaped by the culture, training, and history of the listener. 
Presently, there remain many acoustic parameters related to music performance where the listener's response has not been measured, either in terms of perceptibility or aesthetic response or both. 
For this reason, listener-focused MPA remains one of the most challenging and elusive areas of research. 
However, to the extent that MPA research and its applications depend on perceptual information (e.g., perceived expressiveness), or intend to deliver
perceptually-relevant output (e.g., performance evaluation or reception, similarity ratings), it is imperative to achieve a fuller understanding of the perceptual relevance of the manipulation and interaction of performance characteristics (e.g., tempo, dynamics, articulation). 
The subsequent paragraphs provide a brief overview of the relevant literature on music perception and MPA, along with some discussion of the relevance of this information for current and future work in both MPA and in MIR in general.

\subsection{Musical expression}
When it comes to listener judgments of a performance, it remains poorly understood which aspects are most important, salient, or pertinent for the listener's sense of satisfaction.
According to \citet{schubert_taxonomy_2014}, listeners are very concerned with the   notion of `expressiveness' which is a complex, multifaceted construct.
Performance expression is commonly defined as `variations in musical parameters by a singer or instrumentalist' \citep{dibben_understanding_2014}. 
In other words, performance expression implies the intentional application of systematic variation on the part of the performer. On the other hand, expressive performance (or `expressiveness') implies a judgement (either implicit or explicit) on the part of a listener.

As stated by \citet{devaney_inter-versus_2016}, however, not all variation is expressive: ``The challenge [...] is determining which deviations are intentional, which are due to random variation, and which are due to specific physical constraints that a given performer faces, such as bio-mechanical limitations [...]. In regard
to physical limitations, these deviations may be both systematic and observable in collected performance data, but may not be perceptible to listeners.''
Thus, identifying the variation in a performance that would be \textit{intended as} expressive is only the first step.  
Discovering which performance characteristics contribute to an expressive performance requires dissecting what listeners deem `expressive' as well as understanding the relation and potential differences between measured and perceived performance features.

For instance, expressiveness is genre and style dependent, meaning that the perceived \textit{appropriate} level and style of expression in a pop ballad will be
different from a jazz ballad, and that expression in a Baroque piece will be different from that of a Romantic piece~---~something that has     been referred to as
`stylishness' \citep{fabian_baroque_2009, kendall_communication_1990}.
For example, the timing difference between the primary melody and the accompaniment tends to be wider in jazz than in classical music, and there is evidence that the direction of difference is reversed, i.e., the melody leads the accompaniment in classical piano music \citep{goebl_melody_2001, palmer_assignment_1996} while it follows in the case of jazz \citep{ashley_nt_2002}.
Similarly, syncopation created by anticipating the beat is normative in pop genres but appears
to be reversed in jazz music where syncopation is created by delaying the onset of the melody \citep{dibben_understanding_2014}.

In addition to style-related expression, there is the perceived \textit{amount} of expressiveness, which is considered independent  of stylishness \citep{schubert_dimensions_2006}.
Finally, \citet{schubert_taxonomy_2014} distinguish a third `layer' of expressiveness, \textit{emotional expressiveness}, which arises from a        performer's manipulation of   various features specifically to
alter or enhance emotion.
This is distinct from \textit{musical} expressiveness, or expressive variation, which more generally refers to the manipulation   of compositional elements by the performer in order to be `expressive' without necessarily needing to express a specific emotion. 
Practically speaking, however, it may be difficult for listeners to separate these varieties of expressiveness  \cite[p.293]{schubert_taxonomy_2014}, and research has
demonstrated that there are interactions between them \citep[e.g.,][]{vieillard_expressiveness_2012}.

\subsection{Expressive variation} \label{ssec:expr_var}
Several scholars have made significant advances in our understanding of the role of timing, tempo, and dynamic variation on listeners’ perception of music. 
As noted in Section~\ref{sec:descript}, the subtle variations in tempo and dynamics executed by a performer have been shown to play a large role in highlighting and segmenting musical structure. 
For instance, the perception of metrical structure is largely mediated through changes in timing and articulation within small structural units such as the measure, beat, or sub-beat,
whereas the perception of formal structures are largely communicated through changes across larger segments such as phrases 
\citep[e.g.,][]{sloboda_communication_1983, gabrielsson_once_1987, palmer_assignment_1996, behne_musikpsychologische_1993}. 
An experiment by \citet{sloboda_communication_1983} found that listeners were      better able to identify the meter of an ambiguous passage when performed by a more
experienced performer.
This suggests that even subtle changes in articulation and timing---more easily executed by an expert performer---play an important role in communicating structural
information to the listener.
Through measuring the differences in the performers' expressive variations, Sloboda identified dynamics and articulation~---in particular, a \textsl{tenuto} articulation---~as the most important  features for communicating which notes were accented.

The extent to which a listener's musical expectations align with a performer's expressive variations appears an important consideration. 
For example, because of the predictable relation between timing   and structural segmentation, it has been demonstrated that listeners find it difficult to detect
timing (and duration) deviations from a `metronomic' performance when the pattern and placement of those deviations are stylistically typical \citep{repp_patterns_1990,repp_constraint_1992,ohriner_grouping_2012}.
Likewise, \citet{clarke_imitating_1993} found pianists able to more accurately reproduce a    performance when the timing profile was `normative' with regards to the musical structure, and also found listeners' aesthetic judgments to be highest for those performances with the original timing profiles       compared with those that were inverted or altered.
    
In addition to communicating structural information to the listener, performance features such as timing and dynamics have also been studied extensively for
their role in contributing to a perceived `expressive' performance \citep[see][]{clarke_rhythm_1998, gabrielsson_performance_1999}. 
For instance, a factor analysis by \citet{schubert_taxonomy_2014} examined the features and qualities that may be related to perceived expressiveness, finding that dynamics had the highest impact on the factor labeled
`emotional expressiveness.' 
Recent work by \citet{battcock_acoustically_2019} showed attack rate to be the most important predictor of intensity (or ``arousal'' in terms of two-dimensional models of emotion). 
While this work was not strictly performance analysis since the authors measured elements that correspond to fixed directions from a score (e.g., mode; pitch height), the authors do analyze attack rate, which is related to timing. 
Specifically, the authors point out that understanding the role of timing is confounded by the fact that it encompasses several distinct musical properties such as tempo and rhythm. 
Although the authors do not attempt to segregate these phenomena (tempo and rhythm) in their perceptual experiments, it is clear that for a performer, adjusting the
tempo (globally or locally) would influence the attack rate, and therefore have an impact on perceived intensity. 

The relation between changes in various expressive parameters and their effect on perceived tension ratings has been fairly well studied but with conflicting results.
\citet{krumhansl_perceptual_1996} found that in an experiment comparing an original performance to versions with flat dynamics, flat tempo, or both,
listeners' continuous tension ratings were not affected, implying that tension was primarily conveyed by the melodic, harmonic, and durational elements central to the composition (rather than the performance). 
A similar result was reported by \citet{farbood_interpreting_2013} where repetitions of the same verse across a single performance~---as well as a harmonic reduction of it---~were found to produce strongly correlated tension ratings.
However, \citet{gingras_linking_2016} studied the relation between musical structure, expressive variation, and listeners' ratings of musical tension, and found that variations in expressive timing were most predictive of listeners' tension ratings.

It is equally important to empirically test assumptions about the perceptual effects of expressive variation. 
For instance, some aspects of so-called `micro-timing' variation~---~defined as small, systematic, intentional deviations in timing~---~have been debated with regard to their perceptual effects. 
In particular, micro-timing has been suggested as one of the principle contributors to the perception of `groove' \citep{iyer_embodied_2002, roholt_groove_2014}.
In fact, there is a sizable portion of literature dedicated to this phenomenon, and the role of micro-timing in generating embodied cognitive responses \citep{dibben_understanding_2014}. 
However, \citet{davies_effect_2013} parametrically varied the amount of micro-timing in certain jazz, funk, and samba rhythm patterns, and, contrary to popular belief, found that systematic micro-timing generally led to decreased ratings of perceived groove, naturalness, and liking. 
Similarly, \citet{fruhauf_music_2013} found that the highest ratings of perceived `groove quality' were given to drum patterns that were perfectly quantized, and that increasing systematic micro-timing (by shifting either forwards or backwards), resulted in lower quality ratings.

While the role of expressive variation in timbre and intonation has generally been less studied, there has been substantial attention given to the expressive qualities
of the singing voice, where these parameters are especially relevant  \citep[see][]{sundberg_singing_2018}. 
For instance, \citet{sundberg_intonation_2013}, found that a sharpened intonation at a phrase climax contributed to increased perception of expressiveness and excitement, and \citet{siegwart_acoustic_1995} found that listener preferences   were correlated with certain spectral components such as the relative strength of the fundamental and the value of the spectral centroid.
Similarly, the role of ornamentation in contributing to perceived expression, skill, or overall quality, has been largely overlooked, especially as it relates
to music outside of the classical canon. Some exceptions include research showing subjective preferences for an idealized pitch contour and timing profile of the
Indian classical music ornament \textit{Gamak} \citep{gupta_objective_2012}, 
and, in pop music, the expressive and emotional effects of portamento (or pitch `slides'), as well as the so-called `noisy' sounds of the voice have been
theorized to be of strong importance in generating an emotional response \citep{dibben_understanding_2014}.
In the latter case, no actual perceptual experiments have been conducted to investigate this claim, however, it is consistent with ethological research on the role of vocalizations and sub-vocalizations in affective communication \citep{huron_affect_2015}.

The reason why expressive variation is so enjoyable for listeners remains largely an open research question.
Expressive variation is assumed to be the most important cue to a listener that they are hearing a uniquely \textit{human} performance and is
regularly hailed as the key component in communicating an aesthetically pleasing performance. 
As mentioned above, its role appears to go beyond bolstering the communication of musical structure. 
And, as pointed out by Repp, even a computerized or metronomic performance will contain grouping cues \citep{repp_obligatory_1998}. 
However, one prominent theory suggests that systematic performance deviations (such as tempo) may generate aesthetically pleasing expressive performances in part due to
their exhibiting characteristics that mimic `natural motion' in the physical world \citep{gjerdingen_shape_1988, todd_dynamics_1992, repp_music_1993,
todd_kinematics_1995, van_noorden_resonance_1999} or human movements or gestures \citep{ohriner_grouping_2012, broze_iii_animacy_2013}.
For instance, \citet{friberg_does_1999}, suggested that the shape of final \textsl{ritardandi} matched the velocity of runners coming to a stop and \citet{juslin_five_2003} includes `motion principles' in his model of performance expression.

   \subsection{Mapping and Predicting Listener Judgments}
In order to isolate listeners' perception of features that are strictly performance-related, several    scholars have investigated listeners' judgments across multiple
performances of the same excerpt of music \citep[e.g.,][]{repp_patterns_1990,  fabian_musical_2008}. 
A less-common technique relies on synthesized constructions or manipulations of performances, typically   using some kind of rule-based system to manipulate certain musical parameters 
    \citep[e.g.,][]{repp_expressive_1989, sundberg_how_1993, clarke_imitating_1993, repp_obligatory_1998}, and       frequently making use of continuous data
collection measures \citep[e.g.,][]{schubert_taxonomy_2014}.
    
From these studies, it appears that listeners (especially `trained' listeners) are capable not only   of identifying performance characteristics such as phrasing,
    articulation, and vibrato, but that they are frequently able to identify them in a manner that is aligned with the performer's intentions \citep[e.g.,][]{nakamura_communication_1987, fabian_baroque_2009}.

    However, while listeners may be able to identify performers' intentions, they may not have the perceptual acuity to identify certain features with the same precision allowed by acoustic measures. For instance, a study by  \citet{howes_relationship_2004} showed there was no correlation between measured and perceived vibrato onset times. 
Similarly, \citet{geringer_continuous_1995} found that listeners consistently identified increases in intensity (crescendos) with a greater perceived magnitude of contrast than
the decreases in intensity (decrescendos) regardless of the actual magnitude of change. 
This suggests that there are some measurable performance parameters that may not map well to human perception.
For example, an objectively measurable difference between a `deadpan' and `expressive' performance does not necessarily translate to \textit{perceived expressivity}, especially if the changes in measured performance parameters are structurally normative, as discussed in Section~\ref{ssec:expr_var}. 
Two related papers, by \citet{li_analysis_2015} and \citet{sulem_perception-based_2019}, describe research attempting to better understand the communication chain from score interpretation to performance and
performance to perception, respectively. 
The former attempted to match quantitative acoustic measures with expressive musical terms (commonly used in score directions as the principal means of
communicating expressive instruction), while the latter asked performers to match
the same expressive musical terms in terms of their perceived emotion along a common dimensional model of emotion \citep[i.e.,][]{russell_circumplex_1980}.
This work lays the foundation for future research to empirically examine the full chain of communication; in attempting to manipulate the same acoustic measurements
it may be possible to predict perceived musical and emotional correlates. 
    
An important but rarely discussed consideration is the relation between observed differences in a model and the perceptual evaluation of those differences by a listener. 
For instance, \citet{dixon_perceptual_2006} experiment with various methods for extracting perceived tempo information in relation to expressively performed excerpts with an emphasis on some of the assumptions of beat-tracking algorithms.
They discuss the presumption that what is desirable in a beat-tracking model is typically to accurately mark what was performed rather than what was perceived, even though the two may differ. 
In particular, they note that the perceived beat is smoother than the performance data would indicate.
\citet{busse_toward_2002} evaluated expert listener judgments of the optimal `swing style' of performed jazz piano melodies that were either unmodified, or modified according to one of four `derived'
models with note durations, onset positions, and velocities determined by performers' averages, or else modified by one of three simple `mechanical' models with the same parameters fixed by simple ratio relationships.
Unsurprisingly, the unaltered and derived models were generally preferred to the mechanical models. 
(It is well known that \textit{some} randomness is required in order for a performance to sound convincingly human, and various jitter functions have been implemented in computer music software for this reason since the 1980s.)
Despite that one might predict human preference for one of the original (performed) melodies, several of the derived models were not rated statistically different from the original melodies, suggesting that the averaged parameter values created a realistic model. 
However, the parameters of the unmodified originals were not reported nor compared against each other or those of the derived models, making it impossible to examine any difference thresholds across the measured parameters in terms of their impact on the swing ratings.   
\citet{devaney_inter-versus_2016} used a `singer identification' task to explore differences between inter-singer variability and intra-singer similarity across different performance parameters using both classification and listening experiments.
In general, listeners performed the singer identification task better than chance but far below the abilities of the computational model.
However, there were some similarities between the model parameters and the features reported by listeners as important determinants of their classification (e.g.,
vibrato, pitch stability, timbre, breathiness, intonation). 
Furthermore, the same pair of singers that `confused' the model were the same two conflated by listener judgments.
These experiments represent excellent examples from a scarce pool of research attempting to bridge MIR and cognitive approaches to performance research. 
However, only a comparison of systematic manipulations between
contrived stimuli will allow sufficient control over the individual parameters necessary to make definitive conclusions about the perceptibility of variations in performance and their contribution to the aesthetic value of a performance. 

Given a weak relation between a measured parameter and listeners' perception of that parameter, another important question arises: is the parameter itself not useful in modeling human perception, or is the metric simply inappropriate?
For example, there are many aspects of music perception that are known to be categorical (e.g., pitch)  in which case a continuous metric would not work well in a model designed to predict human ratings.

Similarly, there is the consideration of the role of the \textit{representation} and transformation of a measured parameter for predicting perceptual ratings.
This question was raised by \citet{timmers_predicting_2005}, who examined the representation of tempo and dynamics that best predicted listener judgments of musical similarity.
This study found that, while most existing models rely on normalized variations of tempo and dynamics, the absolute tempo and the interaction of tempo and loudness were better predictors.

Finally, there are performance features that are either not captured in the audio signal or else not represented in a music performance analysis that may well contribute to a listener's perception. 
For instance, if judgements of perception are made in a live setting, then many visual cues~---such as performer movement, facial expression, or attire---~will be capable of altering the listener's perception
\citep{huang_what_2011,juchniewicz_influence_2008,wapnick_effects_2009,livingstone_facial_2009,silvey_role_2012}.
Importantly, visual information such as performer gesture and movement may contribute to embodied sensorimotor engagement, which is thought to be an essential component
of music perception \citep[e.g.,][]{leman_role_2014, bishop_performers_2018}, and could therefore be influential on ratings of performance aesthetics and/or musical expression. 

Clearly, the execution of multiple performance parameters is important for the perception of both
small-scale and large-scale musical structures, and appears to have
a large influence over listeners' perception and experience of the emotional and expressive aspects of a performance.

Since the latter appears to carry great significance for both MPA and music perception research, it suggests that future work ought to focus on disentangling the relative weighting of the various features controlled by performers that contribute to an expressive performance. 
Since it is frequently alluded to that a performer's manipulation of musical tension is one of the strongest contributors to an expressive performance, further
empirical research must attempt to systematically break down the concept of tension as a high-level feature into meaningful collections of smaller, well-defined features that would be useful for MPA.

\subsection{Challenges}
The research surveyed in this section highlights the importance of human perception in MPA research, especially as it pertains to the communication of emotion, musical structure, and creating an aesthetically pleasing performance.
In fact, the successful modeling of perceptually relevant performance attributes, such as those that mark `expressiveness,' could have a large impact not only for MPA
but for many other areas of MIR research, such as computer-generated performance, automatic accompaniment, virtual instrument design and control, or
robotic instruments and HCI (see, for example, the range of topics discussed by \citet{kirke_guide_2013}).
A major obstacle impeding research in this area is the inability to successfully isolate (and therefore understand) the various performance characteristics that contribute to a so-called `expressive' performance from a listener's perspective. 
Existing literature reviews on the topic of MPA have not been able to shed much light on this problem, in part because researchers frequently disagree on (or conflate) the
various definitions of `expressive,' or else findings appear inconsistent across the research, likely as a result of different methodologies, types of
comparisons, or data.  
As noted by \citet{devaney_inter-versus_2016}, combining computational and listening experiments could lead to a better understanding of \textit{which} aspects of variation are important to observe and model.
Careful experimental design and/or meta-analyses across both MPA and cognition research, as well as cross-collaboration between MIR and music cognition researchers, may
therefore prove fruitful endeavors for future research.

\section{Performance Assessment}\label{sec:assessment}
    Assessment of musical performances deals with providing a rating of a music performance with regard to specific aspects of the performance such as accuracy, expressivity, and virtuosity.
    Performance assessment is a critical and ubiquitous aspect of music pedagogy: students rely on regular feedback from teachers to learn and improve skills, recitals are used to monitor progress, and selection into ensembles is managed through competitive auditions.
    The performance parameters on which these assessments are based are not only subjective but also ill-defined, leading to large differences in subjective opinion among music educators \citep{thompson_evaluating_2003,wesolowski_examining_2016}.
    However, other studies have shown that humans tend to rate prototypical (average) performances higher than individual performances \citep{repp_aesthetic_1997,wolf_tendency_2018}. This might indicate that performances are rated based on some form of perceived distance from an `ideal' performance.
    Apart from music education, assessment of performances is also an important area of focus for the evaluation of computer generated music performances \citep{bresin_evaluation_2013} where researchers have primarily focused on listening studies to understand the effect of musical knowledge and biases on rating performances \citep{poli_role_2014} and the degree to which computer generated performances stack up against those by humans \citep{schubert_algorithms_2017}.

    Work within assessment-focused MPA deals with modeling how humans assess a musical performance. The goal is to increase the objectivity of performance assessments \citep{mcpherson_assessing_1998} and to build accessible and reliable tools for automatic assessment. While this might be considered a subset of listener-focused MPA, its importance to MPA research and music education warrants a tailored review of research in this area.
    
    Over the last decade, several researchers have worked towards developing tools capable of automatic music performance assessment. These can be loosely categorized based on:
    \begin{inparaenum}[(i)] 
        \item the parameters of the performance that are assessed, and
        \item the technique/method used to design these systems.
    \end{inparaenum}
    
    \subsection{Assessment parameters}
        Tools for performance assessment evaluate one or more performance parameters typically related to the accuracy of the performance in terms of pitch and timing \citep{wu_towards_2016,vidwans_objective_2017,pati_assessment_2018,luo_detection_2015}, or quality of sound (timbre) \citep{knight_potential_2011,romani_picas_real-time_2015,narang_acoustic_2017}.  In building an assessment tool, the choice of parameters may depend on the proficiency level of the performer being assessed. For example, beginners will benefit more from feedback in terms of low-level parameters such as pitch or rhythmic accuracy as opposed to feedback on higher-level parameters such as articulation or expression. Assessment parameters can also be specific to culture or the musical style under consideration, for example, in the case of Indian classical music the nature of pitch transitions or \textit{gamakas} plays an important role \citep{gupta_objective_2012}, while correct pronunciation of syllables is a strict requirement for Chinese \textit{Jingju} music \citep{gong_automatic_2018}.

        Assessment tools can also vary based on the granularity of assessments. Tools may simply classify a performance as `good' or `bad' \citep{knight_potential_2011, nakano_automatic_2006}, or grade it on a scale, say from $1$ to $10$ \citep{pati_assessment_2018}. Systems may provide fine-grained note-by-note assessments \citep{romani_picas_real-time_2015,schramm_automatic_2015} or analyze entire performances and report a single assessment score \citep{nakano_automatic_2006,pati_assessment_2018,huang_automatic_2019}.
    
    \subsection{Assessment methods}
        While different methods have been used to create performance assessment tools, the common approach has been to use descriptive features extracted from the audio recording of a performance, based on which a classifier predicts the assessment. This approach requires availability of performance data (recordings) along with human (expert) assessments for the rated performance parameters. 

        The level of sophistication of classifiers was limited especially for early attempts, in which classifiers such as Support Vector Machines were used to predict human ratings. In these systems, the  descriptive features became an important aspect of the system design. In some approaches, standard spectral and temporal features such as spectral centroid, spectral flux, and zero-crossing rate were used \citep{knight_potential_2011}. In others, features aimed at capturing certain aspects of music perception were hand-designed using either musical intuition or expert knowledge \citep{nakano_automatic_2006,abeser_automatic_2014,romani_picas_real-time_2015,li_analysis_2015}. For instance, \citet{nakano_automatic_2006} used features measuring pitch stability and vibrato as inputs to a simple classifier to rate the quality of vocal performances. Several studies also attempted to combine low-level audio features with hand-designed feature sets \citep{luo_detection_2015,wu_towards_2016,vidwans_objective_2017}, as well as incorporating information from the musical score or reference performance recordings into the feature computation process \citep{devaney_study_2012,mayor_performance_2009,vidwans_objective_2017,bozkurt_dataset_2017,molina_fundamental_2013,falcao_dataset_2019}. 
    
        Recent methods, however, have transitioned towards using advanced machine learning techniques such as sparse coding \citep{han_hierarchical_2014,wu_learned_2018,wu_assessment_2018} and deep learning \citep{pati_assessment_2018}. Contrary to earlier methods which focused on hand-designing musically important features, these techniques input raw data (usually in the form of pitch contours or spectrograms) and train the models to automatically learn meaningful features so as to accurately predict the assessment ratings.  
        
        \begin{figure*}[t]
            \centerline{
            \includegraphics[width=\textwidth]{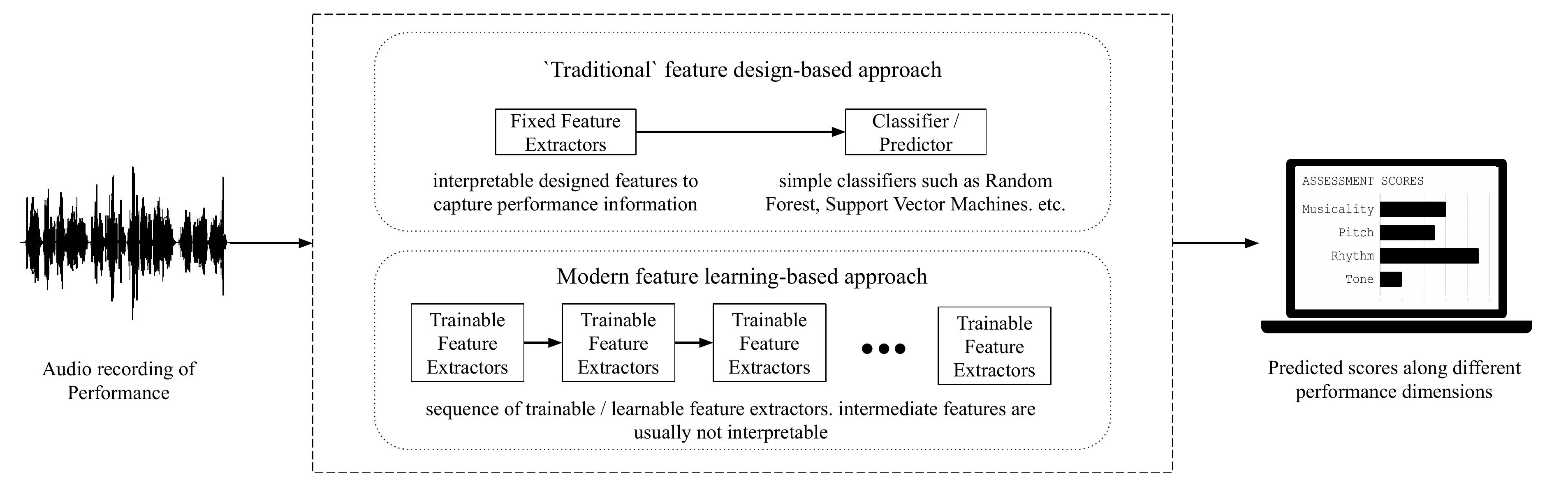}}
            \caption{Schematic showing the comparison between different approaches for music performance assessment}
        \label{fig:mpa_assessment}
        \end{figure*}

        In some ways, this evolution in methodology has mirrored that of other MIR tasks: there has been a gradual transition from feature design to feature learning (compare Figure~\ref{fig:mpa_assessment}). Feature design and feature learning have an inherent trade-off. Learned features extract relevant information from data which might not be represented in the hand-crafted feature set. This is evident from their superior performance at assessment modeling tasks \citep{wu_assessment_2018,pati_assessment_2018}. However, this superior performance comes at the cost of low interpretability. Learned features tend to be abstract and cannot be easily understood. Custom-designed features, on the other hand, typically either measure a simple low-level characteristic of the audio signal or link to high-level semantic concepts such as pitch or rhythm which are intuitively interpretable. Thus, such models allow analysis that can aid in the interpretation of semantic concepts for music performance assessment. For instance, \citet{gururani_analysis_2018} analyzed the impact of different features on an assessment prediction task and found that
        \begin{inparaitem}[]
            \item features measuring tempo variations were particularly critical, and that 
            \item score-aligned features performed better than score-independent features.
        \end{inparaitem}
           
    \subsection{Challenges}
        In spite of several attempts across varied performance parameters using different methods, the important features for assessing music performances remain unclear. This is evident from the average accuracy of these tools in modeling human judgments. Most of the presented models either work well only for very select data \citep{knight_potential_2011} or have comparably low prediction accuracies \citep{vidwans_objective_2017, wu_towards_2016}, rendering them unusable in most practical scenarios {\citep{eremenko_performance_2020}}. While this may be partially attributed to the subjective nature of the task itself, there are several other factors which have limited the improvement of these tools. 
        First, most of the models are trained on small task-specific or instrument-specific datasets that might not reflect noisy real-world data. This reduces the generalizability of these models. The problem becomes more serious for data-hungry methods such as deep learning which require large amounts of data for training. The larger datasets ($>3000$ performances) based on real-world data are either not publicly available (for example, the FBA dataset \citep{pati_assessment_2018}) or only provide intermediate representations such as pitch contours (for example, the MAST melody dataset \citep{bozkurt_dataset_2017}). Thus, more efforts are needed towards creating and releasing larger performance datasets for the research community. Second, the distribution of ground-truth (expert) ratings given by human judges is in many datasets skewed towards a particular class or value \citep{gururani_analysis_2018}. This makes it challenging to train unbiased models. Finally, the performance parameters required to adequately model a performance are not well understood. While the typical approach is to train different models for different parameters, this approach necessitates availability of performance data along with expert assessments for all these parameters. In many occasions, such assessments are either not available or are costly to obtain. For instance, while the MAST rhythm dataset \citep{falcao_dataset_2019} contains performance recordings (and pass/fail assessment ratings) for around $1000$ students, the finely annotated (on a $4$-point scale) version of the same dataset contains only $80$ performances.  An interesting direction for future research might consider leveraging models which are successful at assessing a few parameters (and/or instruments) to improve the performance of models for other parameters (and/or instruments). This approach, usually referred to as transfer learning, has been found to be successful in other MIR tasks \citep{choi_transfer_2017}.
        
    In addition to the data-related challenges, there are several other challenging problems for MIR researchers interested in this domain. Better techniques need to be developed to factor the score (or reference) information into the assessments. So far, this has been accomplished by either using dynamic time warping (DTW) based methods \citep{vidwans_objective_2017,bozkurt_dataset_2017,molina_fundamental_2013} to compute distance-based features between the reference and the performance or by computing vector similarity between features extracted from the performance and the reference \citep{falcao_dataset_2019}. However, expressive performances are supposed to deviate from the score and simple distance-based features may fail to adequately capture the nuances. The problem of how to incorporate this information into the assessment computation process remains an open problem.

    Another area which requires attention from researchers lies in improving the ability to interpret and understand the features learned by end-to-end models. This will play an important role in improving assessment tools. Interpretability of neural networks is still an active area of research in MIR, and performance assessment is an excellent testbed for developing such methods. 

\section{Conclusion}\label{sec:apps}
    The previous sections outlined insights gained by MPA at the intersection of audio content analysis, empirical musicology, and music perception research.  These insights are of importance for better understanding the process of  making music as well as affective user reactions to music. 

    \subsection{Applications}
    The better understanding of music performance enables a considerable range of applications spanning a multitude of different areas including systematic musicology, music education, MIR, and computational creativity, leading to a new generation of music discovery and recommendation systems, and generative music systems.

    The most obvious application example connecting MPA and MIR is music tutoring software. Such software aims at supplementing 
        teachers by providing students with insights and interactive feedback by analyzing and assessing the audio of practice sessions. The ultimate goals of an interactive music tutor are to highlight problematic parts of the student's performance, provide a concise yet easily understandable analysis, give specific and understandable feedback on how to improve, and individualize the curriculum depending on the student's mistakes and general progress. Various (commercial) solutions are already available, exhibiting a similar set of goals. These systems adopt different approaches, ranging from traditional music classroom settings to games targeting a playful learning experience. Examples for tutoring applications are SmartMusic,\footnote{MakeMusic, Inc., \href{https://www.smartmusic.com}{www.smartmusic.com}, last accessed 04/11/2019} Yousician,\footnote{Yousician Oy, \href{https://www.yousician.com}{www.yousician.com}, last accessed 04/11/2019} Music Prodigy,\footnote{The Way of H, Inc., \href{http://www.musicprodigy.com}{www.musicprodigy.com}, last accessed 04/11/2019} and SingStar.\footnote{Sony Interactive Entertainment Europe, \href{http://www.singstar.com}{www.singstar.com}, last accessed 04/11/2019}
    However, many of these tools are not reliable enough to be used in educational settings. More studies are needed to properly evaluate the usability of performance assessment systems in real classroom environments \citep{eremenko_performance_2020}.

    Performance parameters have a long history being either explicitly or implicitly part of MIR systems. For instance, core MIR tasks such as music genre classification and music recommendation systems have a long history of utilizing tempo and dynamics features successfully \citep{fu_survey_2011}. 
   
    Another area which has relied extensively on using performance data is the field of generative modeling. Much of the recent research has been on generating expressive performances with or without a musical score as input. While the vast majority of this body of work has focused on piano performances \citep{cancino-chacon_basis_2016,malik_neural_2017,jeong_graph_2019,jeong_virtuosonet_2019,oore_this_2020,maezawa_rendering_2019}, there are a few studies focused on other instruments such as violin and flute \citep{wang_performancenet_2019}. The common thread across these approaches is that they use end-to-end data-driven techniques to generate the performance (either predict note-wise performance features such as timing, tempo and dynamics, or directly generate the audio) given the score as input. While these methods have achieved some success, they mostly operate as black boxes, and hence, lack in their ability to either provide deeper insights regarding the performance generation process or exert any form of explicit control over different performance parameters. There have been some attempts to alleviate these limitations. For instance, \citet{maezawa_rendering_2019} tried to learn an abstract representation capturing the musical interpretation of the performer. This could allow generation of different performances of the same piece with varying interpretations. More studies like this would allow better modeling of musical performances and improving the quality and usability of performance generation systems. 
 
    \subsection{Challenges}
    Despite such practical applications, there are still many open topics and challenges that need to be addressed. The main challenges of MPA have been summarized at
    the end of each of the previous sections. The related challenges to the MIR community, however, are multi-faceted as well.
    First, the fact that the majority of the presented studies use manual annotations instead of automated methods should encourage the MIR community to re-evaluate the
    measures of success of their proposed systems if, as it appears to be, the outputs lack the robustness or accuracy required for a detailed analysis even for
    tasks considered to be `solved.' 
    Second, the missing separation of composition and performance parameters when framing research questions or problem definitions can impact not only interpretability and reusability of insights but might also reduce algorithm performance. If, for example, a music emotion recognition system does not differentiate between the impact of core musical ideas and performance characteristics, it will have a harder time differentiating relevant and irrelevant information. Thus, it is essential for MIR systems to not only differentiate between score and performance parameters in the system design phase but also analyze their respective contributions during evaluation.
    Third, when examining phenomena that are complex and at times ambiguous~---such as `expressiveness'---~it is imperative to fully define the scope of the associated terminology.  Inconsistently used or poorly defined terms can obfuscate results making it more challenging to build on prior work or to propagate knowledge across disciplines. 
   Fourth, a greater flow of communication between MIR and music perception communities would bolster research in both areas. However, differing methodologies, tools, terminology, and approaches have often created a barrier to such an exchange \citep{aucouturier_mel_2012}.  One way of facilitating this communication between disciplines is to maximize the interpretability and reusability of results. In particular, acknowledging or addressing the perceptual relevance of predictor variables or results, or even explicitly pointing to a possible gap in the perceptual literature, can aid knowledge transfer by pointing to `meaningful' or perceptually-relevant features to focus subsequent empirical work. 
   In addition, it would be prudent to ensure that any underlying assumptions of perceptual validity (linked to methods or results) are made overt and, where possible, supported with empirical results.
    Fifth, lack of data continues to be a challenge for both MIR core tasks and MPA; a focus on approaches for limited data \citep{mcfee_software_2015}, weakly labeled data, and unlabeled data \citep{wu_labeled_2018} could help address this problem. There is, however, a slow but steady growth in the number of datasets available for performance analysis, indicating growing awareness and interest in this topic.
Table~\ref{table:dataset} lists the most relevant currently available datasets for music performance research. Note that 22 of the 30 datasets listed have been released in the last 5 years.
        
    In conclusion, the fields of MIR and MPA each depend on the advances in the other field. 
    In addition, music perception and cognition, while not a traditional topic within MIR, can be seen as an important linchpin for the advancement of MIR systems that depend on reliable and diverse perceptual data. Cross-disciplinary approaches to MPA bridging methodologies and data from music cognition and MIR are likely to be most influential for future research.
    Empirical, descriptive research driven by advanced audio analysis is necessary to extend our understanding of music and its perception, which in turn will allow us to create better systems for music analysis, music understanding, and music creation.
    
\begin{landscape}
\begin{table}
\begin{tabular*}{\linewidth}{p{.12\linewidth}p{.18\linewidth}p{.12\linewidth}p{.08\linewidth}p{.12\linewidth}p{.124\linewidth}p{.124\linewidth}}
\textbf{name} & \textbf{reference} & \textbf{instruments} & \textbf{genre} & \textbf{data} & \textbf{size} & \textbf{perf.\ params}\\ \hline\hline

\href{https://archive.org/details/Automatic_Practice_Logging}{APL}  & \cite{winters_automatic_2016} & piano & classical & audio & 621 recordings & piano practice\\
\href{https://zenodo.org/record/3250223}{CBFdataset}                & \cite{wang_adaptive_2019}     & bamboo flute & chinese & audio & 1GB & playing techniques\\
\href{http://www.crestmuse.jp/pedb/}{CrestMusePEDB}                 & \cite{hashida_new_2008}       & piano & classical & xml & 121 performances & timing, dynamics\\
\href{https://zenodo.org/record/2649950}{CSD}                       & \cite{cuesta_analysis_2018}   & vocals & classical & audio, f0 series & 48 recordings & intonation\\
\href{https://zenodo.org/search?page=1&size=20&q=smule&keywords=Smule}{DAMP}& ---                           & vocals & popular & audio & 24874 recordings (14 songs) & singing\\
\href{https://github.com/cwu307/DrumPtDataset}{DrumPT}              & \cite{wu_drum_2016}           & drums & popular & audio & 30 recordings & playing techniques\\
\href{https://www.cs.cmu.edu/~gxia/data/}{Duet}                     & \cite{xia_duet_2015}          & piano & classical & MIDI & 105 performances & timing, dynamics\\
\href{https://www.upf.edu/web/mtg/eep}{EEP}                         & \cite{marchini_sense_2014}    & string quartet & classical & audio & 23 recordings & timing, gestures, bowing techniques \\
\href{https://www.audiolabs-erlangen.de/resources/MIR/2019-GeorgianMusic-Erkomaishvili}{Erkomaishvili} & \cite{rosenzweig_Erkomaishvili_2020} & vocals & Georgian & audio, f0 series, MusicXML & 116 recordings & timing, pitch\\
\href{https://magenta.tensorflow.org/datasets/groove}{Groove MIDI}  & \cite{gillick_learning_2019}  & drums & popular & MIDI & 13.6 hours & drum timing\\
\href{http://mac.citi.sinica.edu.tw/GuitarTranscription/}{GPT}      & \cite{su_sparse_2014}         & guitar & popular & audio & 6580 recordings & playing techniques\\
\href{https://www.idmt.fraunhofer.de/en/business_units/m2d/smt/bass.html}{IDMT-SMT-Bass} & \cite{abeser_feature-based_2010} & bass & popular & audio & 3.6 hours & playing techniques\\
\href{https://www.idmt.fraunhofer.de/en/business_units/m2d/smt/guitar.html}{IDMT-SMT-Guitar} & \cite{kehling_automatic_2014} & guitar & popular & audio & 4700 note events & playing techniques\\
\href{https://ccrma.stanford.edu/damp/}{Intonation}                 & \cite{wager_intonation_2019}  & vocals & popular & audio, f0 series& 4702 performances & singing\\
\href{https://zenodo.org/record/832736}{Jingju-Pitch}               & \cite{gong_pitch_2016}        & vocals & Beijing Opera & f0 series& 13MB & intonation\\
\href{http://www.cp.jku.at/resources/2019_RLScoFo_TISMIR/}{JKU-ScoFo} & \cite{henkel_score_2019}    & piano & classical & audio, MIDI & 16 performances & timing, dynamics\\
\href{http://yannbayle.fr/karamir/kara1k.php}{Kara1k}               & \cite{bayle_kara1k_2017}      & vocals & popular & audio & 1000 songs & singing\\
\href{https://magenta.tensorflow.org/datasets/maestro}{Maestro}     & \cite{hawthorne_enabling_2019}& piano & classical & audio, MIDI & 200 hours & timing, dynamics\\
\href{https://github.com/barisbozkurt/MASTmelody_dataset}{MASTmelody} & \cite{bozkurt_dataset_2017} & vocals & --- & f0 series & 1018 recordings & pass/fail ratings \\
\href{https://zenodo.org/record/2620357}{MASTrhythm}                & \cite{falcao_dataset_2019}    & percussion & --- & audio & 3721 recordings & pass/fail ratings \\
\href{http://mazurka.org.uk}{Mazurka}                               & \cite{sapp_comparative_2007}  & piano & classical & beat markers & 2732 recordings & tempo, dynamics \\
\href{https://gitlab.doc.gold.ac.uk/expressive-musical-gestures/dataset/-/tree/master/piano}{PGD} & \cite{sarasua_datasets_2017} & piano & classical & audio, video, MIDI & 210 recordings & gestures, intentions\\
\href{https://www.upf.edu/web/mtg/quartet-dataset}{QUARTET}         & \cite{papiotis_computational_2016} & string quartet & classical & audio, video & 96 recordings & timing, gestures, bowing techniques\\
\href{http://resources.mpi-inf.mpg.de/SMD/SMD_MIDI-Audio-Piano-Music.html}{SMD} & \cite{muller_saarland_2011} & piano & classical & audio, MIDI & 50 performances & timing, dynamics\\
\href{https://supra.stanford.edu/download/}{SUPRA}                  & \cite{shi_supra_2019}         & piano & classical & piano rolls, MIDI & 478 performances & gestures, timing, dynamics\\
\href{http://www2.ece.rochester.edu/projects/air/projects/URMP.html}{URMP} & \cite{li_creating_2019} & multi & classical & audio, video & 44 pieces & timing, dynamics\\
\href{https://gitlab.doc.gold.ac.uk/expressive-musical-gestures/dataset/-/tree/master/violin}{VGD} & \cite{sarasua_datasets_2017} & violin & classical & audio, EMG, IMU & 960 recordings & position data, playing techniques\\
\href{https://repo.mdw.ac.at/projects/IWK/the_vienna_4x22_piano_corpus/index.html}{Vienna 4x22} & \cite{goebl_numerisch-klassifikatorische_1999} & piano & classical & audio, MIDI & 4 pieces, 22 pianists & timing, dynamics\\
\href{https://zenodo.org/record/1442513#.W7OaFBNKjx4}{VocalSet}     & \cite{wilkins_vocalset_2018}  & vocals & popular & audio & 6GB & singing techniques\\
\href{https://jazzomat.hfm-weimar.de/dbformat/dboverview.html}{WJazzD} & \cite{pfleiderer_inside_2017} & wind instruments & jazz & MIDI & 456 solos & timing, pitch\\

\end{tabular*}
\caption{A list of datasets useful for various tasks within MPA.}
\label{table:dataset}
\end{table}
\end{landscape}

\bibliography{musicperformanceanalysis}

\end{document}